\documentclass[twocolumn,amsmath,amssymb,aps,superscriptaddress]{revtex4-2}

\usepackage{amsmath,amssymb}  
\usepackage{mathtools}        
\usepackage{graphicx}
\usepackage{braket}
\usepackage{xcolor}

\begin{document}
\title{Large-Scale Quantum Device Benchmarking via \\ LXEB with Particle-Number-Conserving Random Quantum Circuits}

\author{Takumi Kaneda}%
\email{u007243b@ecs.osaka-u.ac.jp
}
\affiliation{%
  Graduate School of Engineering Science, The University of Osaka, 1-3 Machikaneyama, Toyonaka, Osaka 560-8531, Japan
}
\author{Keisuke Fujii}%
\affiliation{%
  Graduate School of Engineering Science, The University of Osaka, 1-3 Machikaneyama, Toyonaka, Osaka 560-8531, Japan
}
\affiliation{%
  Center for Quantum Information and Quantum Biology,
  The University of Osaka, 1-2 Machikaneyama, Toyonaka 560-0043, Japan
}%
\affiliation{%
  RIKEN Center for Quantum Computing (RQC),
  Hirosawa 2-1, Wako, Saitama 351-0198, Japan
}%
\author{Hiroshi Ueda}%
\affiliation{%
  Center for Quantum Information and Quantum Biology,
  The University of Osaka, 1-2 Machikaneyama, Toyonaka 560-0043, Japan
}%
\affiliation{
  RIKEN Center for Computational Science (R-CCS),
  Kobe, Hyogo 650-0047, Japan
}
\date{\today}
\begin{abstract}
Linear cross-entropy benchmarking (LXEB) with random quantum circuits is a standard method for evaluating quantum computers. However, LXEB requires classically simulating the ideal output distribution of a given quantum circuit with high numerical precision, which becomes infeasible beyond approximately 50 qubits, even on state-of-the-art supercomputers. As a result, LXEB cannot be directly applied to evaluate large-scale quantum devices, which now exceed 100 qubits and continue to grow rapidly in size.
To address this limitation, we introduce a constraint known as \textit{particle-number conservation} into the random quantum circuits used for benchmarking. This restriction significantly reduces the size of the Hilbert space for a fixed particle number, enabling classical simulations of circuits with over 100 qubits when the particle number is $O(1)$.
Furthermore, we propose a modified version of LXEB, called \textit{MLXEB}, which enables fidelity estimation under particle-number-conserving dynamics. Through numerical simulations, we investigate the conditions under which MLXEB provides accurate fidelity estimates.
\end{abstract}

\maketitle

\section{Introduction}
Quantum computers are emerging as next-generation computational devices with the potential to solve problems that are intractable for conventional classical computers~\cite{dalzell2023quantum}.
In recent years, experimental research in quantum computing has advanced rapidly, leading to several notable breakthroughs.
For example, in 2019, Google reported the first demonstration of so-called quantum supremacy~\cite{arute2019quantum} by showing that a quantum processor could sample from the output distribution of a random quantum circuit faster than the most powerful classical supercomputers at that time.
Subsequent studies have extended these demonstrations to even larger superconducting devices: Google’s 67-qubit processor has been shown to achieve quantum supremacy in sampling tasks \cite{morvan2024phase}, and USTC’s Zuchongzhi 3.0 platform—with 83 qubits—has similarly pushed the boundary of classically intractable sampling \cite{gao2025establishing}.
In the context of random-circuit sampling, these results indicate that quantum processors are venturing into regimes that exceed the capabilities of the most powerful classical simulators.
Moreover, even quantum hardware other than superconducting architectures has reached comparable scales: Quantinuum’s 56-qubit trapped-ion system has recently executed complex circuits with high fidelity, underscoring the growing stability and scalability of ion-trap platforms \cite{decross2024computational}.


However, the practical use of current quantum computers—commonly referred to as Noisy Intermediate-Scale Quantum (NISQ) devices—continues to face numerous challenges.
One major issue is that quantum computers are inherently more susceptible to noise than classical computers, which makes it difficult to obtain error-free and accurate computational results.
To mitigate the effects of such noise and to ensure reliable quantum computation, quantum error correction (QEC) techniques are essential.
Notably, significant progress has been made in quantum error correction.
In 2024, it was experimentally demonstrated for the first time that the lifetime of a logical qubit could exceed that of a physical qubit by employing surface codes~\cite{acharya2024quantum}.
Despite this milestone, current QEC methods require between $10^6$ and $10^8$ physical qubits to realize a fault-tolerant quantum computer (FTQC)~\cite{katabarwa2024early}.
In contrast, today’s quantum devices are limited to approximately 156 qubits at most~\cite{abughanem2025ibm}, highlighting a significant gap between the current state of hardware and the requirements for realizing FTQC.

Although the realization of FTQC remains a major technological challenge, a crucial intermediate step toward this goal is to develop a deep understanding of the properties of current NISQ devices~\cite{preskill2018quantum, eisert2020quantum}.
In particular, many recent QEC schemes, such as those based on surface codes, are built on the assumption that errors affecting different qubits and gates are statistically independent~\cite{fowler2012surface,dennis2002topological}.
However, in real devices, errors often exhibit spatial and temporal correlations—such as those arising from gate crosstalk—which can violate these assumptions and potentially degrade the performance of QEC protocols~\cite{zhou2025characterization,kam2024detrimental}.
Therefore, developing benchmarking techniques that can quantitatively capture such correlated errors is fundamentally important for the design of future hardware and the realization of FTQC~\cite{proctor2025benchmarking, erhard2019characterizing, harper2020efficient, magesan2012characterizing}.
To address this issue, various benchmarking methods have been proposed.
Traditional approaches, such as Standard Randomized Benchmarking, can characterize systems up to a few qubits~\cite{knill2008randomized, dankert2009exact, magesan2011scalable, magesan2012characterizing}.
More scalable techniques, including Cycle Benchmarking and its variants, have been developed to assess the performance of larger quantum systems~\cite{erhard2019characterizing, harper2020efficient, flammia2020efficient, harper2021fast, flammia2021pauli, flammia2021averaged}.

However, benchmarking protocols mentioned above are all built upon Clifford gates, which limits their applicability to more general circuits that incorporate non-Clifford operations. 
To address this limitation, we focus on Linear Cross-Entropy Benchmarking (LXEB), a protocol based on Random Circuit Sampling, which was employed in Google's demonstration of quantum supremacy~\cite{boixo2018characterizing, arute2019quantum}.
LXEB offers a significant advantage over Clifford-based methods by enabling the estimation of overall circuit fidelity even for random quantum circuits that include non-Clifford gates~\cite{mullane2007sampling, arute2019quantum}. 
Nonetheless, LXEB also faces several significant challenges.
In particular, it requires the ideal output probability distribution obtained from a noiseless simulation of the circuit, which becomes exponentially more difficult to compute as the number of qubits increases.
As a result, even with state-of-the-art supercomputers, classical simulations are practically limited to circuits with at most around 50 qubits~\cite{arute2019quantum}.
Moreover, Google's original work~\cite{arute2019quantum} reported that the circuit-level fidelity estimated by LXEB matched that predicted from gate-level fidelities, suggesting that correlated errors could be neglected. However, the theoretical justification for this observation remains unclear. 
Given these limitations, several recent studies have proposed extensions to LXEB aimed at improving its accuracy and scalability~\cite{liu2021benchmarking, chen2023linear, cheng2025generalized}.
For instance, Liu et al.~\cite{liu2021benchmarking} demonstrated that LXEB does not overestimate the effects of correlated noise and supports layer-wise fidelity estimation.
They further demonstrated that when correlations between errors are weak, the fidelity predicted from gate-level metrics agrees with that obtained using LXEB.
Chen et al.~\cite{chen2023linear} addressed the scalability issue of classical simulation by restricting circuits to Clifford gates, thereby enabling efficient and scalable benchmarking while retaining certain features of LXEB. 
Additionally, Cheng et al.~\cite{cheng2025generalized} introduced the concept of ergodicity to strengthen the theoretical foundation of LXEB by analyzing its validity under the assumption of unitary 2$t$-designs, thereby providing new insights into the behavior of noisy random circuits and the design of quantum benchmarking protocols.
While each of these previous works represents a significant advancement, there remains no benchmarking method that simultaneously achieves both scalability and compatibility with random quantum circuits containing non-Clifford gates.

To overcome the limitations of existing methods and enable benchmarking of quantum circuits with non-Clifford gates on devices with more than 100 qubits, we propose an approach based on particle-number-conserving random quantum circuits and a modified version of LXEB, which we call MLXEB.
Specifically, the particle-number-conserving random quantum circuits considered in this study are those in which the number of excitations (i.e., qubits in the $\ket{1}$ state) remains constant throughout the computation.
Such conservation laws naturally arise in physical quantum many-body systems with symmetries, such as fermionic systems, and the corresponding circuits are built using gates that preserve particle number—for example, two-qubit gates with U(1) symmetry.
Physical symmetries such as U(1) symmetry have been effectively utilized in quantum computing. 
In particular, in variational quantum algorithms (VQAs) such as the variational quantum eigensolver (VQE), explicitly incorporating symmetries can drastically reduce the dimensionality of the Hilbert space to be explored.
This, in turn, helps mitigate optimization challenges such as the barren plateau problem~\cite{lacroix2023symmetry, barkoutsos2018quantum, meyer2023exploiting, arrazola2022universal, ayeni2023bottom, gard2020efficient, summer2024anomalous}.
However, to the best of our knowledge, no prior work has explored the use of such structured circuits for benchmarking quantum hardware.

Unlike general random quantum circuits, imposing a particle-number conservation constraint restricts the entanglement structure of the output state.
In particular, when the particle number is $O(1)$, computing the ideal output probability distribution becomes significantly easier via classical simulation.
This provides a major advantage in terms of scalability.
MLXEB leverages this feature by benchmarking quantum devices using the cross entropy between the bitstring distribution obtained from the actual hardware and the ideal distribution computed classically under the particle-number-preserving constraint.
By doing so, MLXEB enables experimental fidelity estimation for large-scale quantum systems, which would be infeasible using conventional LXEB due to the prohibitive cost of classical simulation.

In our numerical experiments, we executed noiseless, particle-number-conserving random quantum circuits on a square lattice for systems ranging from $N = 36$ to $196$ qubits.
We confirmed that, with sufficient circuit depth, the output distributions converge to the Porter-Thomas distribution, and the fidelity estimated by MLXEB, $F_{\text{MLXEB}}$, asymptotically approaches the ideal value $F = 1$.
We evaluated MLXEB under two circuit configurations: one in which the two-qubit gate rotation angles were chosen randomly, and another in which they were fixed to specific values.
In both cases, $F_{\text{MLXEB}}$ exhibited behavior consistent with theoretical expectations.
Notably, although it is unclear whether circuits with fixed gate angles form a unitary 2-design, MLXEB still performed effectively, suggesting that the method remains valid even for circuits with limited randomness.
Furthermore, under a depolarizing noise model, we found that $F_{\text{MLXEB}}$ matched the true fidelity when the noise rate was sufficiently low ($\sim 10^{-3}$).
These results collectively demonstrate that introducing physical symmetries—such as particle-number conservation—can significantly extend the range of quantum circuits that are classically simulatable.
Moreover, under this constraint, a benchmarking metric similar to LXEB can still reliably estimate circuit fidelity.
In particular, we have verified that noiseless simulations remain feasible for systems with up to 1000 qubits, suggesting that the applicability of MLXEB may extend well beyond the scale studied in this study.
Leveraging this classical simulability, and considering typical experimental gate fidelities around 99.9\% (corresponding to a noise rate of $\sim 10^{-3}$) and associated gate-count overhead, MLXEB can practically benchmark devices with at least approximately 70 qubits, thereby accessing regimes beyond the reach of conventional LXEB.
By bridging the gap between classical simulability and experimental feasibility, MLXEB offers a practical and theoretically grounded approach to benchmarking near-term quantum devices—enabling systematic fidelity assessments for complex circuits that incorporate non-Clifford operations.

The remainder of this paper is organized as follows.
In Section~\ref{sec:rc_lxeb}, we review the types of random quantum circuits used in LXEB and the fidelity-related quantities that are evaluated.
Section~\ref{sec:pcrc_mlxeb} introduces the particle-number-conserving random quantum circuits required for MLXEB, and defines the evaluation metric for fidelity estimation under this constraint.
In Section~\ref{sec:numerics}, we describe the specific circuit setup and present numerical validation of MLXEB.
Finally, Section~\ref{sec:summary} provides a summary of our findings and discusses future directions.

\section{LXEB Using\\Random Quantum Circuits}\label{sec:rc_lxeb}

\subsection{Properties Required\\for Random Quantum Circuits}\label{sec:random_qc}
Let $\hat{U}$ be a unitary operator representing a quantum circuit acting on an $N$-qubit system with Hilbert space dimension $D = 2^N$ and computational basis $\{\ket{0}, \ket{1}\}^{\otimes N}$. Applying $\hat{U}$ to the initial quantum state $\ket{\psi_0} = \ket{0}^{\otimes N}$ yields the final state $\ket{\psi} = \hat{U} \ket{\psi_0}$.
Upon measuring all qubits of $\ket{\psi}$ in the computational basis, the system collapses to a product state $\ket{x} = \ket{b_N} \otimes \cdots \otimes \ket{b_1}$, where $b_\ell \in \{0,1\}$ denotes the measurement outcome of the $\ell$-th qubit for $\ell \in [1,N]$. We define $x$ as the integer corresponding to the bit string $(b_N b_{N-1} \ldots b_1)$.
In the absence of noise or other imperfections, the probability of obtaining a particular outcome $x$ is given by 
\begin{equation}
    p(x) = \left| \braket{x|\psi} \right|^2.
    \label{def:p(x)}
\end{equation}

Under the assumption of a sufficiently large Hilbert space, if the unitary operator $\hat{U}$ is sampled uniformly at random from the set of all unitary operators defined on that space, the resulting distribution $\mathcal{P}(p)$ of the probability $p$ converges to the Porter-Thomas distribution~\cite{mullane2007sampling}:
\begin{equation}
    \mathcal{P}_{\rm T}(p) = D e^{-D p}.
    \label{def:PTdist}
\end{equation}
In order for random quantum circuits to be used for fidelity estimation in LXEB, the ideal output probability distribution must follow $\mathcal{P}_{\rm T}$.
Therefore, it is essential to verify, within the regime where the classical simulation of the quantum circuit is tractable, that the user-specified circuit structure indeed produces this expected distribution.

\subsection{LXEB}\label{sec:lxeb}
By executing a random quantum circuit on a real quantum device and performing $N_s$ repeated measurements in the computational basis, one obtains a set of bit strings represented as integers, denoted by $\{x_i\}_{1 \leq i \leq N_s}$. 
LXEB evaluates the fidelity using these outcomes according to the following formula:
\begin{equation}
  F_{\text{LXEB}} = \frac{2^N}{N_s} \left( \sum_{i=1}^{N_s} p_{\text{sim}}(x_i) \right) - 1,
  \label{eq:lxeb}
\end{equation}
as originally proposed in~\cite{arute2019quantum}.
Here, $p_{\text{sim}}(x_i)$ denotes the probability of observing the bit string $x_i$ from the ideal output state of the random quantum circuit.
This quantity must be computed in advance using a classical simulator of the quantum circuit. 

An important property of $F_{\text{LXEB}}$ is that
it coincides with the fidelity
\begin{equation}
  \label{def:fidelity}
  F = \left( {\rm tr}\sqrt{\sqrt{\hat{\rho}} \ket{\psi}\bra{\psi} \sqrt{\hat{\rho}} } \right)^2
\end{equation}
in the limit of a large number of samples $N_s \gg 1$, provided that the noise rate is sufficiently low and the output distribution $p(x)$ can be modeled as a linear combination of the Porter-Thomas distribution and the uniform distribution induced by depolarizing noise.
Here, $\hat{\rho}$ denotes the density matrix representing the mixed state resulting from the effect of noise, and $\ket{\psi}$ is the ideal output state in the absence of noise.
Another well-known property of $F_{\text{LXEB}}$ is that, in the absence of noise ($p_{\text{noise}} = 0$), sampling from the output distribution yields $F_{\text{LXEB}} = 1$. 
As $p_{\text{noise}}$ increases, the Porter-Thomas distribution gradually breaks down, and the output distribution approaches the uniform distribution.
In this regime, $F_{\text{LXEB}}$ asymptotically approaches $0$, and no longer coincides with the fidelity, which instead saturates at the lower bound $1/2^N$ due to the completely mixed output state.

\section{MLXEB Using Particle-Conserving Random Quantum Circuits}\label{sec:pcrc_mlxeb}

\subsection{Particle-Conserving Random Quantum Circuits}\label{sec:pcrc}
As a quantity characterizing the computational basis state $\ket{x}$, we define the particle number $n$ as the expectation value 
\begin{equation}
    n = \Braket{x|(N - \hat{Z}_{\rm tot})/2|x},
    \label{def:N-particle}
\end{equation}
where $\hat{Z}_{\rm tot} = \sum_{\ell} \hat{Z}_{\ell}$, and $\hat{Z}_{\ell}$ is the Pauli-$Z$ operator acting on the $\ell$-th qubit.
Then, we also define the set 
\begin{equation}
    G_n = \left\{ x \,\middle|\, \Braket{x|(N - \hat{Z}_{\rm tot})/2|x} = n\right\}.
\end{equation}
to specify the subspace that conserves the particle number.
The number of elements of $G_n$ is given by $D_n = \binom{N}{n}$.
A unitary operator $\hat{U}'$ that preserves particle number can be written as
\begin{equation}
\hat{U}' = \sum_{n=0}^{N} \left( \sum_{x \in G_n} \sum_{x' \in G_n} U_{x,x'} \ket{x}\bra{x'} \right),
\label{def:p-conceving-circuit-unitary}
\end{equation}
and satisfies unitarity within each particle-number-$n$ subspace:
\begin{equation}
\sum_{x' \in G_n} U_{x,x'} U_{x'',x'}^{*} = \delta_{x,x''}, \quad \text{for all } x, x'' \in G_n.
\end{equation}

An advantage of considering $\hat{U}'$ is that, when it acts on a quantum state expressed as a linear combination of computational basis states $\{ \ket{x} \}_{x \in G_n}$ with fixed particle number $n$, the resulting state remains within the same subspace and can also be represented as a linear combination of $\{ \ket{x} \}_{x \in G_n}$.
Therefore, if such a state is chosen as the initial state, the simulation of the corresponding quantum circuit can be performed by tracking a complex vector of dimension $D_n$.
When $n$ remains relatively small, this enables the simulation of quantum circuits with more than 100 qubits on modern classical computers.

To simplify the discussion, we adopt a specific product state $\ket{o_n} \in G_n$ as the initial state, defined as
\begin{equation}
    \ket{o_n} = \prod_{k=1}^{n} \hat{X}_{i_k} \ket{\psi_0},
    \label{def:initial-state}
\end{equation}
where $\{i_k\}$ is a set of integers satisfying $1 \leq i_1 < i_2 < \cdots < i_n \leq N$.
Applying the particle-number-conserving unitary operator $\hat{U}'$ to the initial state $\ket{o_n}$ results in the final state $\ket{\psi_n} = \hat{U}'\ket{o_n}$.
The corresponding noise-free probability distribution is defined as $p_n(x) = |\braket{x|\psi_n}|^2$.
Note that $p_n(x)$ is identically zero for $x \notin G_n$, and it depends not only on the circuit $\hat{U}'$ but also on the choice of the initial configuration $o_n$.
Following the discussion in Section~\ref{sec:random_qc}, when the dimension $D_n$ of the particle-number-$n$ subspace becomes sufficiently large (e.g., in systems with large $N$), and $\hat{U}'$ is sampled uniformly at random from the space of particle-number-conserving unitaries, the probability distribution $p_n$ converges to the particle-number-rescaled Porter-Thomas distribution:
\begin{equation}
    \mathcal{P}_n(p_n) = D_n e^{-D_n p_n},
    \label{def:rescaled-PTdist}
\end{equation}
as shown in Ref.~\cite{liu2024unitary,mitsuhashi2025unitary}.
\subsection{MLXEB}\label{sec:mlxeb}
In this section, we introduce a phenomenological modification of the fidelity estimator $F_{\rm LXEB}$, discussed in Section~\ref{sec:lxeb}, to enable its application to random quantum circuits that conserve particle number.
Specifically, we define the modified linear cross-entropy benchmarking (MLXEB) for a given particle number $n$ as
\begin{equation}
  F_{{\rm MLXEB},n} = \frac{n_s}{N_s} \left[ \frac{ D_n }{ n_s } \left( \sum_{j=1}^{n_s} p_n(x'_j) \right) - 1 \right],
  \label{def:mlxeb}
\end{equation}
where $n_s$ denotes the number of measurement outcomes (out of $N_s$ total samples) that belong to the particle-number-$n$ subspace, and $\{x'_j\}_{1 \leq j \leq n_s}$ is the set of corresponding bitstring indices.
Each $p_n(x'_j)$ represents the ideal (noise-free) probability of observing $x'_j$, as defined in the particle-number-conserving scenario.

When $p_{\rm noise} = 0$, $F_{{\rm MLXEB},n}$ converges to $1$ in the limit of a sufficiently large number of samples $n_s \gg 1$, just as $F_{\rm LXEB}$ does in the noiseless case.
This is because, in the absence of noise, $n_s = N_s$, and the definition of $F_{{\rm MLXEB},n}$ becomes equivalent to that of $F_{\rm LXEB}$ in Eq.~\eqref{eq:lxeb}, with the total Hilbert space dimension $D$ replaced by the subspace dimension $D_n$.
Similarly, the Porter-Thomas distribution $\mathcal{P}$ is replaced by its particle-number-rescaled counterpart $\mathcal{P}_n$ under this constraint.
In the opposite limit of depolarizing noise, the measured bitstrings are sampled from the uniform distribution $p(x) = D^{-1}$.
Assuming $n_s \gg 1$, the summation $\sum_{j=1}^{n_s} p_n(x'_j)$ converges to $N_s / D$, while the fraction $n_s / N_s$ converges to $D_n / D$.
Substituting these into the definition of $F_{{\rm MLXEB},n}$ shows that it asymptotically approaches $0$ in this high-noise regime.

As shown above, $F_{{\rm MLXEB},n}$ reproduces the behavior of $F_{\rm LXEB}$ in both extreme limits.
However, whether it can reliably evaluate circuit fidelity in the intermediate regime remains nontrivial.
As we demonstrate through the numerical experiments presented below, $F_{{\rm MLXEB},n}$ remains a valid estimator when the overall fidelity of the circuit is sufficiently high to rise above the sampling noise, and the depth of the ideal circuit is large enough for the output distribution $p(x)$ to approach the rescaled Porter-Thomas distribution.

\section{Validation by Numerical Simulations}\label{sec:numerics}

\subsection{Construction of Particle-Conserving Random Quantum 
Circuits}\label{sec:setup_RQC}
\begin{figure*}[t]
  \centering
  \includegraphics[width=175mm]{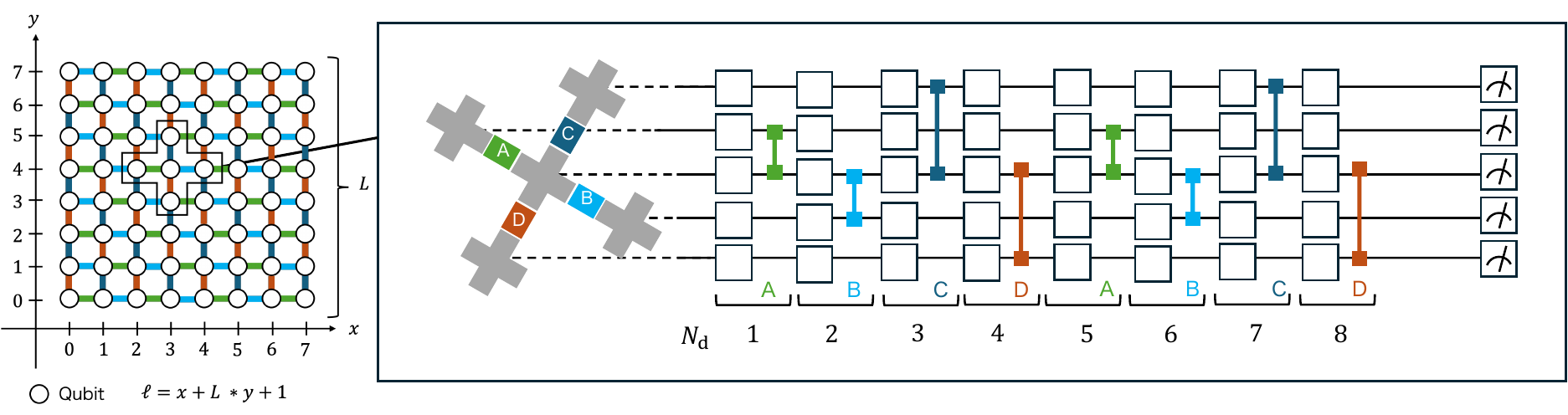}
  \caption{Unit structure of a particle-number-conserving random quantum circuit on a square lattice. A single cycle consists of four layers labeled A, B, C, and D, which together define one unit of circuit depth.}
  \label{fig:pc-circuit}
\end{figure*}

This work focuses on particle-number-conserving random quantum circuits arranged on a two-dimensional square lattice.
We numerically simulate quantum circuits that implement the unitary operator \(\hat{U}'\) , which is composed of layers of two-qubit gates that preserve U(1) symmetry, equivalent to particle number conservation.
Even within this restricted class, it has been shown that the output probability distribution still converges to the Porter-Thomas distribution~\cite{liu2024unitary,mitsuhashi2025unitary}.
An overview of the layer-wise structure of the circuit used in our numerical experiments is shown in Figure~\ref{fig:pc-circuit}.
In this figure, the white square symbols represent single-qubit $R_z$ gates, where each qubit is assigned an independent and uniformly random rotation angle $\theta \in [0, 2\pi)$.
The two-qubit gates, illustrated as colored dumbbell shapes, are XXZ-type gates derived from the canonical decomposition~\cite{Khaneja2001PRA,Kraus2001PRA}, with a shared XY interaction.
The gates are defined by
\begin{equation}
w(\alpha_1,\alpha_2) = \exp\left(i[\alpha_1(X \otimes X + Y \otimes Y) + \alpha_2 Z \otimes Z]\right),
\end{equation}
where $i$ denotes the imaginary unit.

We consider two types of parameter settings for the two-qubit gates: (i) random angles $\alpha_1, \alpha_2 \in [0, 2\pi)$ drawn independently and uniformly, and (ii) fixed angles $\alpha_1 = \pi/8$ and $\alpha_2 = 0$. In the random-angle case, it has been shown that the circuit forms a unitary $t$-design within the fixed particle-number subspace~\cite{liu2024unitary,mitsuhashi2025unitary}, thereby ensuring convergence to the Porter-Thomas distribution.
Recent work by Suzuki et al.~\cite{suzuki2024globalrandomnessrandomlocal} reported that fixing the parameters of two-qubit gates can enhance global randomness in a shallow-depth circuit and accelerate convergence to the Porter-Thomas distribution by better approximating a unitary 2-design~\cite{harrow2009random, dankert2009exact}.
While these results were established in the context of generic random quantum circuits without symmetry constraints, we are particularly interested in U(1)-symmetric random quantum circuits, where such convergence properties have not yet been fully explored.
Motivated by these findings, we hypothesize that introducing fixed-angle two-qubit gates may similarly enhance convergence to the Porter-Thomas distribution even under U(1) symmetry constraints.
To test this hypothesis, we compare random-angle and fixed-angle configurations and examine their impact on the convergence rate.
In the context of MLXEB, achieving convergence with shallow circuit depths is particularly valuable, as it enhances the robustness of the benchmark against noise (i.e., higher $p_{\rm noise}$).
In our circuits, two-qubit gates play the central role in directly generating entanglement and superposition between distinct qubits.
Therefore, when using fixed-angle configurations, the choice of gate parameters becomes critical.
In this study, we adopt $\alpha_1 = \pi/8$ and $\alpha_2 = 0$ to ensure that the two-qubit gates effectively generate entanglement by transforming input states such as $\ket{1}\otimes \ket{0}$ or $\ket{0}\otimes \ket{1}$ into maximally entangled states.
This enables the realization of the Porter-Thomas distribution even with shallow-depth circuits.
However, under the fixed-angle constraint, whether the output distribution still converges to the rescaled Porter-Thomas form remains nontrivial and must therefore be verified numerically.
The scheduling of the XXZ-type two-qubit gates in layers A, B, C, and D follows the same pattern as that used in the original LXEB benchmarking study~\cite{arute2019quantum}.

In the MLXEB protocol for a given particle number $n$, the quantum circuit follows the sequence: (i) applying $n$ single-qubit $X$ gates to construct the initial state $\ket{o_n}$; (ii) performing $N_{\rm cycle}$ cycles of single- and two-qubit gate layers in the order A$\rightarrow$B$\rightarrow$C$\rightarrow$D; and (iii) measuring all qubits in the computational basis.
Depolarizing noise~\cite{nielsen2010quantum} with probability $p_{\rm noise}$ is independently applied to each gate operation and measurement: (i) after each single-qubit; (ii) after each measurement operation; and (iii) after each two-qubit gate. 
Due to the high computational cost of simulating noisy circuits using the density matrix formalism, we instead adopt the stochastic method (also known as the Monte Carlo wavefunction method)~\cite{Molmer1993JOSAB}.
Specifically, one of the Pauli operators $X$, $Y$, or $Z$ is randomly applied with probability $p_{\rm noise}$ after each single-qubit gate and before each measurement operation.
Similarly, after each two-qubit gate, a two-qubit Pauli operator randomly selected from the set $\{I, X, Y, Z\}^{\otimes 2} \setminus \{I \otimes I\}$ is applied with the same probability.
The noise pattern is independently resampled for each of the $N_s$ circuit executions.
Under sufficient sampling, this method is known to yield results that are statistically equivalent to those obtained from full density matrix simulation\cite{arute2019quantum}.
Similarly, the fidelity defined in Eq.~\eqref{def:fidelity} can be statistically estimated by averaging the overlap $\braket{\psi_{\rm noise}|\psi_n}$ between the noisy final state $\ket{\psi_{\rm noise}}$—generated by circuits with inserted Pauli noise gates—and the ideal, noiseless state $\ket{\psi_n}$.

\subsection{Validation of the Porter-Thomas Distribution}\label{sec:numerical_pt}
\begin{figure}[htbp]
  \centering
  \begin{minipage}[t]{0.48\textwidth}
    \raggedright \textbf{(a)}\\[-0em]
    \includegraphics[width=\linewidth]{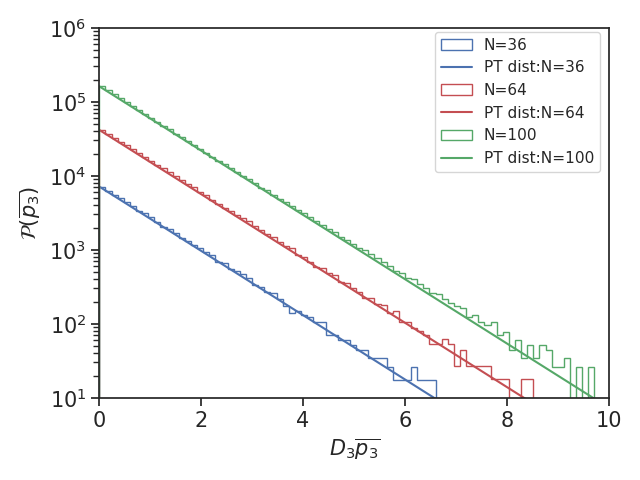}
  \end{minipage}
  \hfill
  \begin{minipage}[t]{0.48\textwidth}
    \raggedright \textbf{(b)}\\[-0em]
    \includegraphics[width=\linewidth]{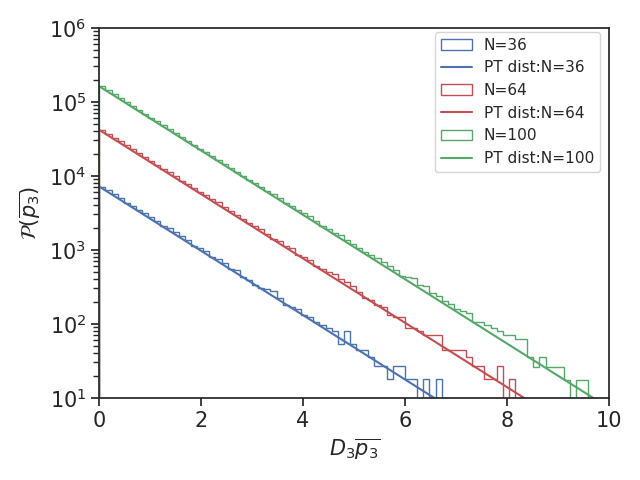}
  \end{minipage}
  \caption{
    Probability distributions of output probabilities of $n=3$ from particle-number-conserving random quantum circuits under two different gate parameter settings:
    (a) two-qubit gate rotation angles $\alpha_1$ and $\alpha_2$ are randomly selected from the interval $[0, 2\pi)$;
    (b) fixed parameters $\alpha_1 = \pi/8$ and $\alpha_2 = 0$.
    The bin width for the histogram is set to $\Delta(D_3 \bar{p}_3) = 0.12$.
    See Section~\ref{sec:numerical_pt} for additional computational conditions.
  }
  \label{fig:pt_dis}
\end{figure}

Figure~\ref{fig:pt_dis} presents results verifying whether the quantum circuits constructed in Section~\ref{sec:setup_RQC} behave as suitable random quantum circuits for MLXEB evaluation under noiseless condition ($p_{\rm noise} = 0$).
We consider system sizes of $N = 6 \times 6$, $8 \times 8$, and $10 \times 10$, each with a fixed particle number of $n = 3$.
The depth per cycle is set to 4, and the total circuit depth is $N_{\rm d} = 160$, which is sufficiently deep relative to the system size. The number of random circuit instances is set to $N_c = 10$.
In Figure~\ref{fig:pt_dis}, the averaged probability distribution $\bar{p}_3$ is computed as the ensemble average over $N_c$ instances, with the probability vector for each instance sorted in ascending order before averaging.
The state-vector simulations for fixed-particle-number subspaces in large-scale systems are performed using a custom Python code built upon internal routines from the quantum spin simulator QS$^3$~\cite{ueda2022quantum,Ueda2023}.
This implementation enables explicit access to the full probability distribution $p_3$ even for systems as large as 100 qubits.

The simulation results confirm that the probability distribution $\mathcal{P}$ is well described by the rescaled Porter-Thomas distribution $\mathcal{P}_n = D_n e^{-D_n p_n}$ for systems with 36 to 100 qubits, regardless of whether the two-qubit gate rotation angles are randomly chosen or fixed.
This agreement holds under sufficiently large circuit depth $N_{\rm d}$, indicating that the output statistics $\mathcal{P}$ converge to the expected Porter-Thomas behavior in both configurations.

\subsection{MLXEB Evaluation in the Absence of Noise and Size Dependence}
\begin{figure}[htbp]
  \centering
  \begin{minipage}[t]{0.48\textwidth}
    \raggedright \textbf{(a)}\\[-0em]
    \includegraphics[width=\linewidth]{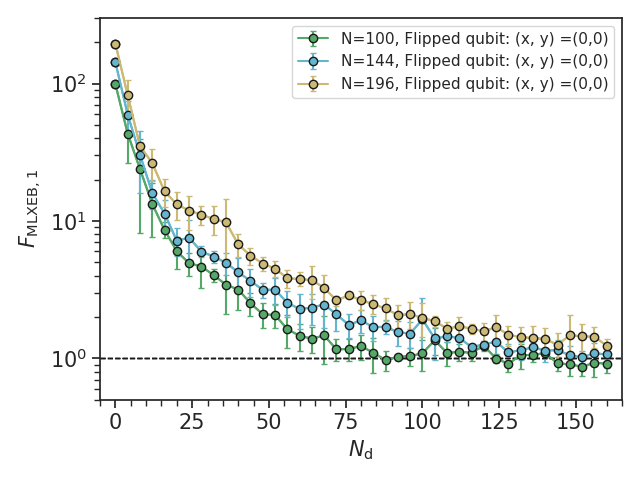}
  \end{minipage}
  \begin{minipage}[t]{0.48\textwidth}
    \raggedright \textbf{(b)}\\[-0em]
    \includegraphics[width=\linewidth]{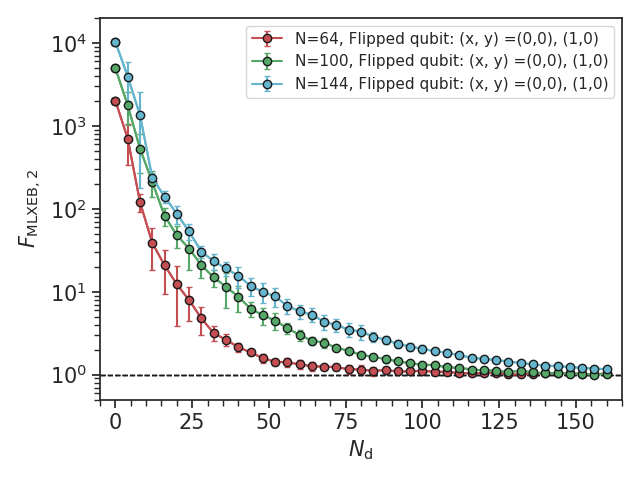}
  \end{minipage}
  \begin{minipage}[t]{0.48\textwidth}
    \raggedright \textbf{(c)}\\[-0em]
    \includegraphics[width=\linewidth]{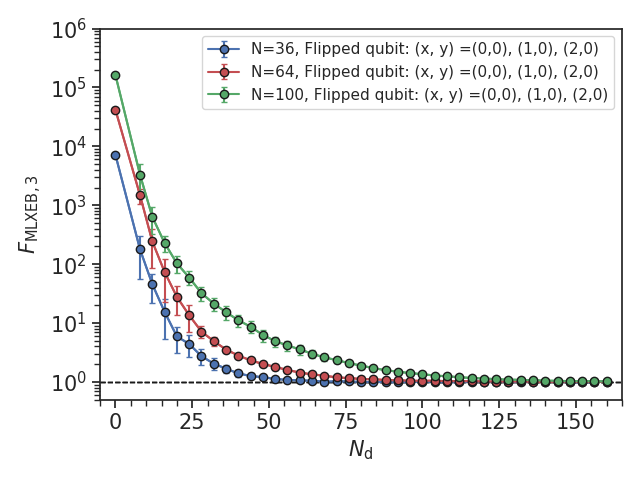}
  \end{minipage}
  \caption{
    Circuit depth dependence of (a) $F_{{\rm MLXEB},1}$, (b) $F_{{\rm MLXEB},2}$, and (c) $F_{{\rm MLXEB},3}$ under noiseless conditions ($p_{\rm noise} = 0$).
    The initial state is prepared by applying $X$ gates to qubits at positions $(0,0), \ldots, (n-1,0)$ of the vacuum state $\ket{\psi_0}$.
    The rotation angles of the two-qubit gates are randomly chosen from the interval $\alpha_1, \alpha_2 \in [0, 2\pi)$.
    The number of random circuit instances and the number of samples per instance are set to $(N_c, N_s) = (5, 100800)$.
    The dashed line indicates the ideal fidelity $F = 1$, which is trivially achieved in the absence of noise.
    }
  \label{fig:mxeb-d-NOD1}
\end{figure}

Having confirmed in Section~\ref{sec:numerical_pt} that the constructed random quantum circuits exhibit the desired statistical properties, we now proceed to evaluate MLXEB. 
In the noiseless case ($p_{\rm noise} = 0$), the modified linear cross-entropy benchmarking value $F_{{\rm MLXEB},n}$ is expected to asymptotically converge to the ideal fidelity $F = 1$ as the total circuit depth $N_{\rm d}$ increases.
Figure~\ref{fig:mxeb-d-NOD1} shows the dependence of $F_{{\rm MLXEB},n}$ on the circuit depth $N_{\rm d}$ for particle numbers $n = 1$, $2$, and $3$ under noiseless conditions.

For all system sizes considered, we observe that $F_{{\rm MLXEB},1}$, $F_{{\rm MLXEB},2}$, and $F_{{\rm MLXEB},3}$ exhibit exponential convergence toward unit fidelity as the circuit depth $N_{\rm d}$ becomes comparable to the number of qubits $N$.
In the shallow-circuit regime (small $N_{\rm d}$), the output quantum state remains insufficiently randomized, and the probability distribution $\mathcal{P}$ deviates from the Porter-Thomas distribution.
Consequently, MLXEB fails to accurately characterize the circuit fidelity in this regime.

\begin{figure}[bp]
  \centering
  \includegraphics[width=\linewidth]{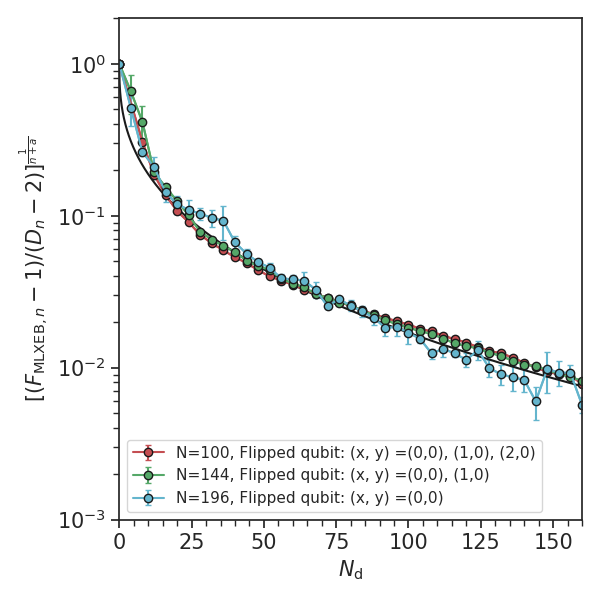}
 \caption{
    Rescaled plot of the data from Figure~\ref{fig:mxeb-d-NOD1}, based on the finite-size scaling relation given in Eq.~\eqref{eq:scaling}.
    The resulting data collapse indicates a universal scaling behavior of $F_{{\rm MLXEB},n}$ across different system sizes and particle numbers.
    }
 \label{fig:Collapse plot}
\end{figure}

To quantitatively examine how the circuit depth $N_{\rm d}$ required for the emergence of asymptotic behavior depends on the system size $N$ and the particle number $n$, we replot the data using the following finite-size scaling relation (Figure~\ref{fig:Collapse plot}):
\begin{equation}
f(N_{\rm d}) = \left[\frac{F_{{\rm MLXEB},n} - 1}{D_n - 2}\right]^{\frac{1}{n + a}}. 
\label{eq:scaling}
\end{equation}
Here, the scaling function $f(N_{\rm d})$ is defined based on the asymptotic behavior of $F_{{\rm MLXEB},n}$:
\begin{equation}
    F_{{\rm MLXEB},n}|_{N_{\rm d}=0}=D_n-1,\quad F_{{\rm MLXEB},n}|_{N_{\rm d} \rightarrow \infty}=1
\end{equation}
From these properties, $f(N_{\rm d})$ must asymptotically approach $1$ as $N_{\rm d} \to 0$, and $0$ as $N_{\rm d} \to \infty$.
The derivation of the scaling function $f(N_{\rm d})$, which is independent of the system size $N$ and the particle number $n$, is based on the following three steps:
\begin{enumerate}
    \item Subtract the trivial asymptotic value 1 from $F_{\rm MLXEB,n}$, to which it converges as $N_{\rm d} \rightarrow \infty$.
    
    \item Normalize $F_{\rm MLXEB,n} - 1$ by dividing by $D_n - 2$, so that the result becomes 1 at $N_{\rm d} = 0$, regardless of $N$ and $n$.
    
    \item Under the condition $n \ll N$, the quantum state in the basis $\{\ket{x} \in G_n\}$ expands to approximately $O(2^n)$ states under the action of a single layer (A, B, C, or D) of the quantum circuit. 
    To eliminate the $n$-dependence of the convergence rate of $F_{\rm MLXEB,n}$ towards $1$, the exponent is divided by a linear function of $n$, specifically $n + a$, where $a$ is a fitting parameter.
\end{enumerate}
Through the above three steps, we expect the behavior of the rescaled function near both limits, $N_{\rm d} = 0$ and $N_{\rm d} \rightarrow \infty$, to be independent of $N$ and $n$.
Whether the scaling function $f(N_{\rm d})$ accurately captures the behavior across the full range of $N_{\rm d}$ must be verified by examining the actual data, particularly in the intermediate-depth regime.

As shown in Figure~\ref{fig:Collapse plot}, when $a=0.3$ was chosen, the behavior of $F_{\text{MLXEB},n}$ for various combinations of $N$ and $n$ is consistently captured by the single scaling function $f(N_{\rm d})$ across the entire range of circuit depths $N_{\rm d}$.
This confirms that the proposed scaling relation accurately captures the fidelity convergence behavior of MLXEB.
Moreover, the overall shape of $f(N_{\rm d})$ suggests that it follows a stretched-exponential form.
To support this, we fit the data for $(N, n) = (100,3)$, $(144,2)$, and $(196,1)$ over the range $N_{\rm d} = 0$–$160$ using the following relation:
\begin{equation}
\left[\frac{F_{{\rm MLXEB},n}-1}{D_n-2}\right]^{\frac{1}{n+a}} = \exp\left[-\left(\frac{N_{\rm d}}{\tau}\right)^{\beta}\right], 
\end{equation}
This yields the fitting parameters $\tau = 2.64\pm0.24$ and $\beta = 0.387\pm0.010$.
This scaling relation is particularly useful for estimating the circuit depth $N_{\rm d}$ required to achieve a desired level of fidelity in MLXEB for larger quantum systems not considered in this study.
It also provides a practical guideline for designing shallow yet reliable benchmarking circuits in large-scale quantum devices.

\begin{figure}[htbp]
  \centering
  \includegraphics[width=80mm]{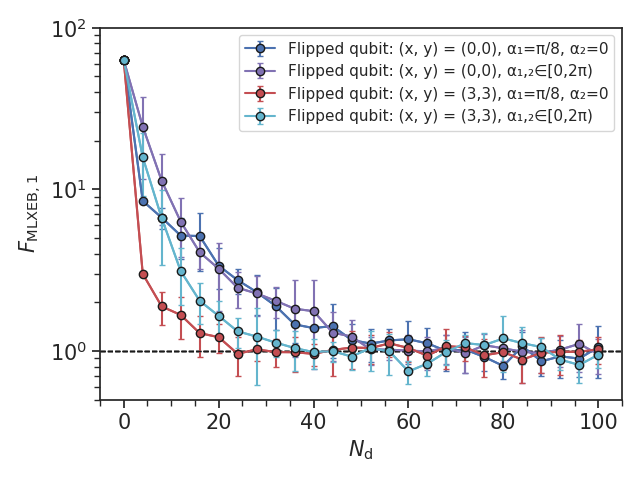}
  \caption{
    Circuit depth dependence of $F_{{\rm MLXEB},1}$ under noiseless conditions ($p_{\rm noise} = 0$) for a system of $N = 64$ qubits.
    Four configurations are compared based on two binary factors: (1) whether the two-qubit gate parameters are fixed ($\alpha_1 = \pi/8$, $\alpha_2 = 0$) or randomly chosen, and (2) whether the initial particle is placed near the center or at the corner of the square lattice.
  }
  \label{fig:compare64}
\end{figure}

Next, we compare the behavior of $F_{\rm MLXEB,1}$ for a system with $N = 64$ and $n = 1$ under four distinct circuit configurations.
Specifically, we consider the following cases: (i) fixed two-qubit gate parameters $\alpha_1 = \pi/8$, $\alpha_2 = 0$; (ii) randomly chosen gate parameters $\alpha_1, \alpha_2 \in [0, 2\pi)$; and for each, we compare the effect of placing the initial particle either near the center of the lattice at position $(\sqrt{N}/2-1, \sqrt{N}/2-1)$ or at the corner $(0, 0)$.
The corresponding results are presented in Figure~\ref{fig:compare64}.
Regardless of whether the two-qubit gate parameters are fixed or randomized, placing the initial particle near the center of the lattice leads to faster convergence of $F_{\rm MLXEB,1}$ to unity at smaller circuit depths $N_{\rm d}$.
This observation is consistent with a qualitative picture based on circuit light cones: the circuit depth required to produce a sufficiently random quantum state—i.e., one in which the state effectively explores the entire Hilbert space—must be at least comparable to the depth needed for the causal light cone, originating from the initial particle position, to cover the entire system.
In this view, initializing the particle near the center allows its light cone to reach all parts of the system more quickly than in the case where the particle starts at the corner, thereby achieving convergence at shallower depths.
In addition, we confirm that fixing the two-qubit gate parameters to $\alpha_1 = \pi/8$ and $\alpha_2 = 0$ accelerates convergence, regardless of the initial particle position.
This trend is consistent with findings reported by Suzuki et al.~\cite{suzuki2024globalrandomnessrandomlocal}, where enhanced global randomness was observed in U(1)-symmetric random circuits by fixing gate parameters.


\subsection{Noise Dependence of MLXEB}\label{sec:noise-dependency}
\begin{figure}[tbp]
  \centering
  \includegraphics[width=80mm]{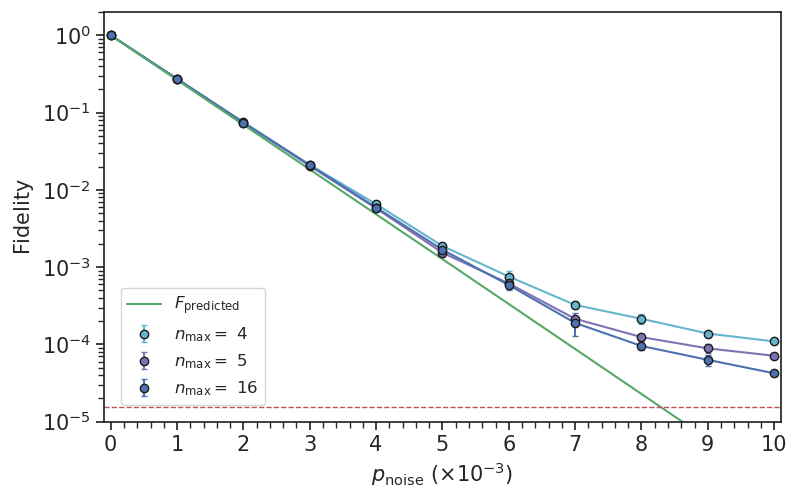}
  \caption{Noise-rate dependence of the fidelity $F$ for a 16-qubit system with circuit depth $N_{\rm d} = 60$. 
           The number of circuit instances and samples per instance are $(N_c, N_s) = (5, 100800)$. 
           The red dashed line indicates the reference value $1/2^N$.}
  \label{fig:mxeb-noise-16}
\end{figure}

We now present the numerical results under noisy conditions for the quantum circuits introduced in Section~\ref{sec:setup_RQC}.
As mentioned in Section~\ref{sec:setup_RQC}, each single-qubit and two-qubit gate is followed by depolarizing noise, and each qubit experiences depolarizing noise just before measurement.
Here, we introduce the error probabilities $e_g$ and $e_q$, which denote the depolarizing noise rates for gate operations and measurements, respectively.
We evaluate the behavior of MLXEB in the perturbative regime, where $e_g, e_q \ll 1$.
The probability that a state with particle number $n$ transitions to another state with a particle number differing by $\delta n$ due to noise is trivially of order $O(p_{\rm noise}^{\delta n})$.
Therefore, to capture the effects of noise up to order $(\delta n+1)$, it is sufficient to retain only the subspaces with particle numbers ranging from $n_{\rm min} = \max[0, n - \delta n]$ to $n_{\rm max} = \min[n + \delta n, N]$.
In this work, we always set $n_{\rm min} = 0$ to improve numerical accuracy.
This is justified because, under the condition $n = O(1) \ll N$, the inequality $\sum_{k=0}^n D_k \ll D_{n+1}$ clearly holds, making the inclusion of lower particle-number sectors computationally inexpensive.
To verify the validity of this truncation strategy, we investigated the noise dependence of the fidelity in a system with $N = 4 \times 4$ qubits and $n = 3$, where the initial particle positions are set to $(0,0)$, $(1,0)$ and $(2,0)$.
We compared three different truncation thresholds: $n_{\rm max} = 4$, $5$, and $16$ (i.e., full Hilbert space).
The results are shown in Fig.~\ref{fig:mxeb-noise-16}.
The two-qubit gate parameters $\alpha_1 \in [0, 2\pi)$ and $\alpha_2 \in [0, 2\pi)$ were independently and uniformly sampled at random.
The green curve represents the lower bound of the predicted circuit fidelity, calculated from the noise rates of all gates and qubits using the following expression:
\begin{equation}
    F_{\text{predicted}} = \prod_{g \in g_1} (1 - e_g) \prod_{g \in g_2} (1 - e_g) \prod_{q \in Q} \left(1 - \frac{2}{3}e_q\right),
\end{equation}
where $g_1$, $g_2$, and $Q$ denote the sets of one-qubit gates, two-qubit gates, and all qubits, respectively. 
Note that the depolarizing noise applied before the measurement induces a bit-flip error with probability $2e_q/3$, which affects the measurement fidelity.
In the numerical experiments, for simplicity, we set all gate error rates $e_g$ and measurement error rates $e_q$ to the same value $p_{\text{noise}}$.
As shown in the figure, the results of truncated simulations with $n_{\rm max} = 4$ and $5$ closely match those from full-vector simulations ($n_{\rm max} = 16$) in the region $p_{\rm noise} \le 5.0 \times 10^{-3}$.
On the other hand, for $p_{\rm noise} \ge 5.0 \times 10^{-3}$, noticeable deviations arise between $F_{\text{predicted}}$ and the fidelities obtained from full-vector simulation.
In particular, for small $N$, the fidelity tends to approach the lower bound $1/2^N$ (shown as the red dashed line) likely due to finite-size effects.

In the remainder of this paper, we focus on the case $n = 1$, $n_{\rm max} = 3$, and $N = 8 \times 8$, which provides a good balance between numerical accuracy and computational cost. 
Furthermore, to reduce the required circuit depth, we fix the two-qubit gate parameters to $\alpha_1 = \pi/8$ and $\alpha_2 = 0$, and place the initial particle at $(\sqrt{N}/2-1, \sqrt{N}/2-1)$ of the square lattice.

\begin{figure}[tbp]
  \centering
  \includegraphics[width=80mm]{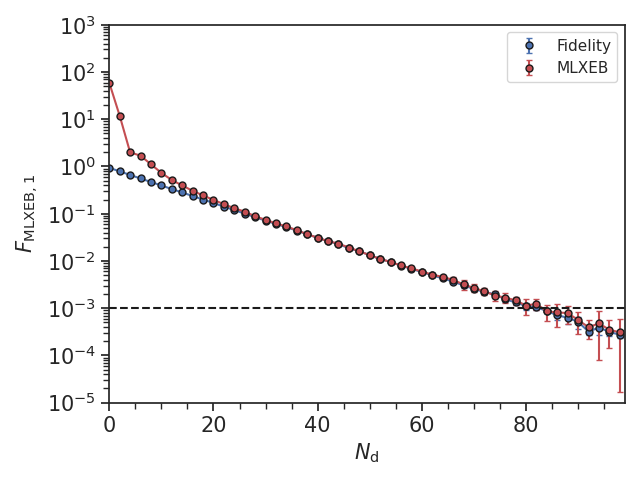}
  \caption{
    Circuit depth dependence of $F_{{\rm MLXEB},1}$ under weak depolarizing noise ($p_{\rm noise} = 0.001$).
    The number of random circuit instances and the number of samples per instance are $(N_c, N_s) = (10, 100800)$, respectively.
  }
  \label{fig:mxeb-cycle-noise}
\end{figure}

We begin by evaluating $F_{\rm MLXEB, 1}$ 
under weak depolarizing noise with $p_{\rm noise} = 10^{-3}$.
The evaluation is performed with accuracy up to third order in $p_{\rm noise}$, and the results are presented in Figure~\ref{fig:mxeb-cycle-noise}.
In this setup, $F_{\rm MLXEB, 1}$ begins to converge to the overall circuit fidelity at circuit depths $N_{\rm d} \gtrsim 20$, consistent with the noiseless simulation shown in Figure~\ref{fig:compare64}.
This result demonstrates that the proposed benchmarking scheme remains robust under perturbative noise.

\begin{figure}[tbp]
  \centering
  \includegraphics[width=80mm]{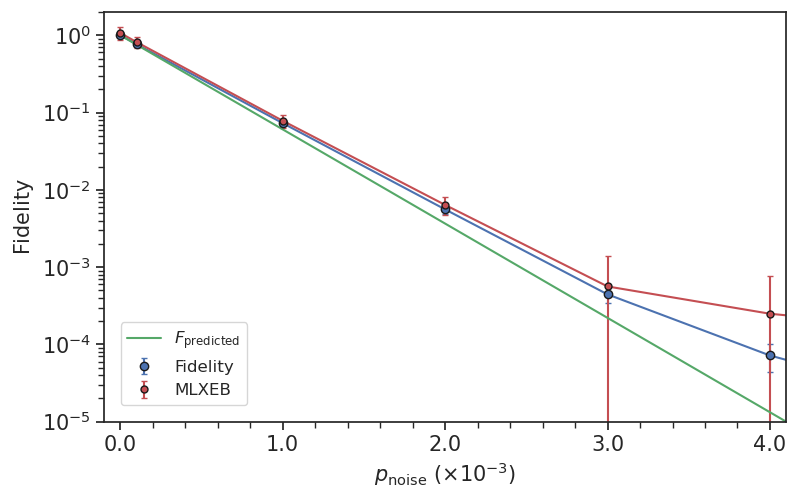}
  \caption{
    Comparison of $F_{{\rm MLXEB},1}$ with the true fidelity for noisy particle-number-conserving random quantum circuits.
    The number of circuit instances and the number of samples per instance are $(N_c, N_s) = (10, 100800)$, respectively.
  }
  \label{fig:mlxeb-noise}
\end{figure}

Finally, we investigate up to what level of noise the MLXEB estimate remains consistent with the actual circuit fidelity.
The results are shown in Fig.~\ref{fig:mlxeb-noise}, where the circuit depth is fixed at $N_{\rm d} = 30$, which is sufficient to ensure convergence to the Porter-Thomas distribution.
Numerical simulations indicate that for the range $1.0 \times 10^{-3} \le p_{\rm noise} \le 2.0 \times 10^{-3}$, the fidelity estimated by MLXEB, $F_{\text{MLXEB},1}$, agrees well with the true circuit fidelity. 
In contrast, for $p_{\rm noise} \ge 3.0 \times 10^{-3}$, a discrepancy emerges between the two.
A likely reason for this deviation is the sampling error, which scales as $O(1/\sqrt{N_s N_c})$~\cite{arute2019quantum}.
This implies that the standard deviation remains non-negligible, and the statistical accuracy is insufficient to reliably resolve the fidelity at higher noise levels. 
In the low-noise regime, we also find that the predicted fidelity $F_{\text{predicted}}$ provides a good approximation of the overall circuit fidelity, validating the analytical expression used in Eq.~(6).

These findings collectively confirm that MLXEB remains robust against realistic levels of depolarizing noise, accurately capturing circuit fidelity up to $p_{\rm noise} \sim 2 \times 10^{-3}$ within the limits of statistical uncertainty.

\section{Conclusion}\label{sec:summary}
In this work, we addressed the limitation that conventional Linear Cross-Entropy Benchmarking (LXEB) is not directly applied to quantum devices with more than 50 qubits due to the high computational cost of exact classical simulation.
To overcome this issue, we introduced the particle number $n$ as a conserved quantity and proposed a modified version of LXEB, referred to as MLXEB.
This method employs particle-number-conserving random quantum circuits and enables estimation of the overall circuit fidelity within a fixed-$n$ subspace.
We numerically validated MLXEB for systems of up to $N = 196$ qubits, assuming a square-lattice interaction topology representative of near-term quantum devices.
We considered two types of random circuits: one with randomly chosen two-qubit gate angles and another with fixed angles.
In both cases, we demonstrated that with sufficient circuit depth, the output probability distribution converges to the rescaled Porter-Thomas distribution defined on the particle-number-$n$ Hilbert subspace.
For the noiseless MLXEB with randomly sampled two-qubit gate parameters, we numerically confirmed that the MLXEB value $F_{\rm MLXEB}$ follows the scaling relation given by Eq.~\eqref{eq:scaling}.
This result allows us to predict, in advance, the circuit depth $N_{\rm d}$ required for accurate fidelity estimation using MLXEB, even for quantum devices larger than the current scale.

On the other hand, when the two-qubit gate parameters are fixed to $\alpha_1 = \pi/8$ and $\alpha_2 = 0$, and the initial particle is placed near the center of the two-dimensional lattice, the output distribution is observed to converge more rapidly to the rescaled Porter-Thomas distribution compared to the case where the gate parameters are randomly selected and the particle is initialized at the lattice corner.
This property is particularly significant when evaluating MLXEB on real quantum hardware, where the number of quantum gates available in a single execution is limited.
However, under this fixed-angle condition, the scaling relation in Eq.~\eqref{eq:scaling} does not appear to hold consistenty across the range of $N$ considered. 
Identifying an appropriate scaling relation for MLXEB under fixed-angle conditions remains an open question for future research.
Furthermore, it remains a theoretical open question to determine which class of unitary $t$-designs these particle-number-conserving random circuits with fixed two-qubit gate parameters belongs to, particularly within the context of U(1)-symmetric ensembles.

In addition, for a system with $N = 64$ qubits, we evaluated MLXEB under perturbative noise using numerically validated simulations.
The results confirmed that MLXEB can estimate the circuit fidelity accurately within the level of statistical uncertainty.


Future work also includes developing advanced MPI-based parallelization schemes and algorithmic improvements to support simulations of larger systems, as well as investigating circuit architectures that are compatible with experimental implementation on real quantum hardware, including constraints such as qubit connectivity, native gate sets, and gate- or qubit-dependent noise characteristics.
By addressing these challenges, MLXEB is expected to become a practical benchmarking method for future quantum devices with $100$ to $1000$ qubits.

This work demonstrates that, under typical experimental noise levels ($p_{\rm noise} \sim 10^{-3}$), MLXEB can provide reliable fidelity estimates for quantum circuits with at least $64$ qubits, and potentially more, depending on the hardware characteristics.

\begin{acknowledgments}
The authors acknowledge helpful discussions and support
by Ryo Watanabe.
This work is partially supported by KAKENHI Grant Numbers JP21H04446, JP21H05182, and JP21H05191 from JSPS of Japan.
We also acknowledge support from MEXT Q-LEAP Grant No. JPMXS0120319794, and from JST COI-NEXT No. JPMJPF2014, and CREST No.JPMJCR24I1 and JPMJCR24I3.
H.U. was supported by the COE research grant in computational science from Hyogo Prefecture and Kobe City through Foundation for Computational Science
\end{acknowledgments}

\bibliographystyle{apsrev4-2}  
\bibliography{cite}
\end{document}